\documentclass[aps,preprint,showpacs,preprintnumbers,amsmath,amssymb]{revtex4}
\usepackage{amsmath,mathrsfs,amsbsy,color,graphicx,bm,amsthm,amsfonts}
\usepackage{units}
\usepackage{bbm}
\usepackage{times}
\usepackage{dcolumn}
\usepackage{mathrsfs}
\usepackage{amsmath,amssymb,epsfig}

\begin{document}

\title{Schwinger effect of Gaussian correlations in constant electric fields}
\author{Shu-Min Wu$^1$}
\author{Hao-Sheng Zeng$^1$}
\email{hszeng@hunnu.edu.cn}
\affiliation{  Department of Physics, Synergetic  Innovation Center for Quantum Effects, and Key Laboratory of Low
Dimensional Quantum Structuresand Quantum Control \\
 of Ministry of Education,
 Hunan Normal University, Changsha, Hunan 410081, China.}


\begin{abstract}
We study the Schwinger effect of Gaussian correlations (quantum entanglement, discord and mutual information) of the continuous-variable two-mode squeezed states shared by Alice and Bob, paying special attention to the difference of the Schwinger effect of correlations between modes of fermion-fermion and qubit-bosonic fields studied previously. We also study the redistribution and conservativeness of the correlations under the Schwinger effect.
\end{abstract}

\vspace*{0.5cm}


\maketitle

\section{Introduction}
In recent years, the topic of relativistic quantum information has received much attention. The behaviours of quantum correlations (entanglement, discord and mutual information) in non-inertial frames \cite{wd10,wd11,Qiang2018,Sun2017,Richter2015} or in the background of black holes \cite{wd12,wd13,zeng2019,Wang2016} have been investigated extensively. In these settings, the Unruh effect or the Hawking effect that involves the production of particle-antiparticle pairs plays the central role for the description of quantum correlations.

Schwinger predicted that the quantum vacuum is unstable under the influence of a strong electromagnetic field and decays into particle-antiparticle pairs\cite{wd17}. The Schwinger effect, in the sense of the production of particle-antiparticle pairs, is similar to the Unruh effect or Hawking radiation, so that a unified form of Bogoliubov transformations that relates the $input$ to the $output$ states exists. However the transformation coefficients have different structures in different effects, which may lead to different influences to the quantum correlations. The authors in references \cite{wd32,wd31} revealed the influence of Schwinger effect to the pairwise quantum correlations of fermion-fermion system and qubit-boson systems respectively. For the sake of completeness, we here want to further study the Schwinger effect of quantum correlations of continuous-variable boson-boson systems. Our main motivation is to reveal the relationship between the Schwinger effects of quantum correlations of boson and fermion fields, and the relationship between Schwinger effect and Unruh effect or Hawking effect of quantum correlations.

Generally speaking, there are various types of continuous variable systems and also various types of states for the given systems. Here we limit our attention to the relevant but specific case of position and momentum continuous spectrum, and choose the two-mode squeezed state as the research object. This type of bosonic state is a typical example of continuous-variable entangled state that approximates to an arbitrarily good extent the EPR pair \cite{wd1}, and especially it is within the reach of current laboratories \cite{wd19}. As will be shown in the text, our research method may be straightforwardly apply to the study of Schwinger effect of other states, such as coherent states and thermal states, as well as Fock states and their simple superposition.

The paper is organized as follows. In Sec. II,  we introduce the measure of quantum entanglement, discord and mutual information for the two-mode Gaussian states, where two different sets of measure are involved. In Sec. III, we present the brief description of Schwinger effect of continuous-variable bosonic fields.
Sec. IV is the main body of our research work, in which the behaviours of various kinds of pairwise quantum correlations (entanglement, discord and mutual information), under the influence of Schwinger effect, are investigated. Finally we give our conclusions and discussions in Sec. V.

\section{Description of correlations for Gaussian states}
A Gaussian state is defined as any state whose characteristic function and quasiprobability distribution are Gaussian functions on the quantum phase space. The properties of a Gaussian state are thus completely specified by the first and second statistical moments of the quadrature operators. As the first moments can be arbitrarily adjusted by marginal displacements, which leave all informationally relevant properties (such as entropy or any measure of correlations) invariant, we can simply set them to be zero. Therefore, the second moments become the unique elements for describing Gaussian states. Combined with our research, we consider a two-mode Gaussian state $\rho_{AB}$ shared by Alice and Bob. We introduce a vector of quadrature operators as $\hat R = ( \hat{X}_A, \hat{P}_A,  \hat{X}_B, \hat{P}_B)^{\sf T}$, where $\hat{X}_i$ and $\hat{P}_i$ ($i=A,B)$ are the position and momentum of each subsystem. The components of $R$ satisfy the canonical commutation relations $[{{{\hat R}_k},{{\hat R}_l}} ] = i{\Omega _{kl}}$ with $\Omega = {{\ 0\ \ 1}\choose{-1\ 0}}^{\oplus{2}}$ being the symplectic matrix. Now we can build a covariance matrix (CM) $\sigma$ with elements ${\sigma _{ij}} = {\rm Tr}\big[ {{{\{ {{{\hat R}_i},{{\hat R}_j}} \}}_ + }\ {\rho _{AB}}} \big]$. For a physically legitimate Gaussian state, its CM $\sigma_{AB}$ must fulfill the uncertainty relation
${\sigma _{AB}} + i\Omega \ge 0$.
The CM of any two-mode Gaussian state can always be put into a block form \cite{wd23,wd24,wd25}
\begin{equation}\label{w1}
 \sigma_{AB}= \left(
                      \begin{array}{cc}
                        \mathcal{A} & \mathcal{X} \\
                        \mathcal{X}^{\rm T} & \mathcal{B} \\
                      \end{array}
                    \right),
\end{equation}
where the sub-matrices have the forms $\mathcal{A}={\rm diag}(a,a)$, $\mathcal{B}={\rm diag}(b,b)$, and $\mathcal{X}={\rm diag}(c_1,c_2)$.
The symplectic invariants are defined by  $J_1=a^2$, $J_2=b^2$, $J_3=c_1c_2$ and $J_4=
\det \sigma_{AB}=(ab-c_1^2)(ab-c_2^2)$.

In this paper, we will study the Schwinger effect of quantum entanglement, discord and mutual information of the two-mode Gaussian states. There are several methods for measuring these correlations, and we will adopt two metrics for our comparative study. The first is the most widely used set of entanglement logarithmic negativity, discord and mutual information based on the von Neumann entropy. The intention by this set of measure is to make comparisons with the results of previous researches for fermion-fermion or qubit-boson systems. The second set of measure may be called the ``R\'{e}nyi-2 measure of correlations", because all the measures of correlations are based on the concept of R\'{e}nyi-2 entropy. It is introduced for making a comparison between the two sets of measure of correlations.

\subsection{The first set of measure of correlation}

In this set of measure, the entanglement is measured by the most widely used logarithmic negativity which quantifies the violation of the positivity of the partial transposition of a density matrix \cite{Peres1996}. This measure has the advantage of computability for high-dimensional or even continuous-variable systems. For the two-mode Gaussian state $\sigma_{AB}$ expressed in Eq.(\ref{w1}), the logarithmic negativity that quantifies the degree of quantum entanglement between the two modes can be rewritten as \cite{wd25}
\begin{equation}\label{w2}
N({\sigma_{AB}})=
\max\bigg\{0,\,- \log{\mu}_-\bigg\}\,,
\end{equation}
where $\mu_-$ is the symplectic eigenvalue of the partial transpose of $\sigma_{AB}$ and is given by $2{\mu}_{-}^2=\delta-\sqrt{\delta^2-4\det \sigma_{AB}}$ with $\delta=J_1+J_2-2J_3$.

The mutual information describes the total correlation between subsystems A and B, including quantum and classical correlations. The mutual information based on the von Neumann entropy is defined as $I(\rho_{AB})=S(\rho_{A})+S(\rho_{B})-S(\rho_{AB})$ with $S(\rho)$ the von Neumann entropy of quantum state $\rho$. For the two-mode Gaussian state with symplectic $\sigma_{AB}$ in Eq.(\ref{w1}), the mutual information may be calculated through expression\cite{wd27},
\begin{equation}\label{w4}
I({\sigma_{AB}})=f(a)+f(b)-f({\nu}_-)-f({\nu}_+),
\end{equation}
where the function is defined as $f(x)=[\frac{x+1}{2}] \log[{\frac{x+1}{2}}]- [\frac{x-1}{2}] \log[{\frac{x-1}{2}}]$, and
$2{\nu}_{\mp}^2=\Delta \mp\sqrt{\Delta^2-4\det \sigma_{AB}}$ with $\Delta=J_1+J_2+2J_3$ are the symplectic eigenvalues of CM $\sigma_{AB}$.

Quantum discord is defined as the difference of mutual information and classical correlation, i.e., $D({\sigma_{AB}})=I({\sigma_{AB}})-C({\sigma_{AB}})$. Unlike entanglement, quantum discord is asymmetrical with respect to the two subsystems. The ``quantum $A$ discord"  based on the Alice's conditional information obtained by Bob's measurements has the form
for a two-mode Gaussian state \cite{wd25,wd26}
\begin{equation}\label{w3}
D({\sigma_{A|B}})=f(b)-f({\nu}_-)-f({\nu}_+)+f(\varepsilon),
\end{equation}
with parameter $\varepsilon$ defined by $\varepsilon=\frac{a+ab+c_1 c_2}{b+1}$. With the exchanging of $a\rightarrow b$, one can pass from the $A$ discord to the $B$ discord.

\subsection{The second set of measure of correlations}
R\'{e}nyi-$\alpha$ entropy is a useful quantity in quantum science. We can use it to construct the measure of correlations for both discrete and continuous variable systems. R\'{e}nyi-$\alpha$ entropy is defined as
\begin{equation}\label{Q1}
S_\alpha(\rho)=\frac{1}{1-\alpha}\ln {\rm tr}(\rho^\alpha),
\end{equation}
which reproduces the von Neumann entropy in the limit $\alpha\rightarrow1$. For the Gaussian states, the special R\'{e}nyi-2 entropy can be  computed very easily
\begin{equation}\label{Q2}
S_2(\rho)=-\ln {\rm tr}(\rho^2)=\frac{1}{2}\ln(\det \sigma),
\end{equation}
with $\sigma$ the CM of the Gaussian state with density matrix $\rho$.

Following the definition of entanglement of formation \cite{Wootters1996}, by replacing the entanglement monotone with the R\'{e}nyi-2 entropy, a measure of entanglement for a Gaussian state with covariance matrix $\sigma_{AB}$ is constructed\cite{ZZ1},
\begin{equation}\label{ent}
    N_2({\sigma_{AB}})=\inf_{\gamma_{AB}}\frac{1}{2}\ln(\det\gamma_{A}),
\end{equation}
where $\gamma_{A}$ is the marginal CM of subsystem $A$ by partially tracing $\sigma_{AB}$ over the modes of subsystem $B$, and the minimization is over all pure Gaussian states with CM $\gamma_{AB}$ that fulfil $0<\gamma_{AB}\leq\sigma_{AB}$ and $\det\gamma_{AB}=1$. [For two real symmetric matrices $M$ and $N$, the statement $M\geq N$ means that the matrix $M-N$ has all non-negative eigenvalues.] For the two-mode Gaussian states with CM given by Eq.(\ref{w1}), this R\'{e}nyi-2 entanglement can be reduced to a more compact form \cite{ZZ3,ZZ4}
\begin{equation}\label{Q3}
N_2({\sigma_{AB}})=\frac{1}{2}\ln(\inf_{\theta\in[0,2\pi]}  m_\theta ),
\end{equation}
with
$$
\begin{array}{c}
 m_\theta = 1+\bigg[c_1(ab-c_2^2)-c_2+\cos\theta\sqrt{[a-b(ab-c_2^2)][b-a(ab-c_2^2)]} \bigg]^2 \\
 \times \bigg\{2(ab-c_2^2)(a^2+b^2+2c_1c_2)+\sin\theta(a^2-b^2)\sqrt{1-\frac{[c_1(ab-c_2)+c_2]^2}{[a-b(ab-c_2^2)][b-a(ab-c_2^2)]}} \\
 - \frac{\cos\theta[2abc_2^3+(a^2+b^2)c_1c_2^2+((1-2b^2)a^2+b^2)c_2-ab(a^2+b^2-2)c_1]}{\sqrt{[a-b(ab-c_2^2)][b-a(ab-c_2^2)]}}\bigg\}^{-1}.
\end{array}
$$

By replacing the von Neumann entropy with R\'{e}nyi-2 entropy in the definition of mutual information and discord of Eqs.(\ref{w4}) and (\ref{w3}), one obtains the corresponding measures of correlations based on the R\'{e}nyi-2 entropy.
For the two-mode Gaussian states with CM given by Eq.(\ref{w1}), the R\'{e}nyi-2 ``quantum $A$ discord" $D_2({\sigma_{A|B}})$ has the expression \cite{wd27,ZZ1,ZZ3}
\begin{equation}\label{Q4}
D_2({\sigma_{A|B}})=\ln(b)- \frac{1}{2}\ln(\det\sigma_{AB})+\frac{1}{2}\ln(\gamma),
\end{equation}
where $\gamma=[2|c_1c_2|\sqrt{G}+G
+c_1^2c_2^2](b^2-1)^{-2}$ with$G=(ab^2-a-bc_2^2)(ab^2-a-bc_1^2)$ for $[ab^2c_2^2-c_1^2(a+bc_2^2)][ab^2c_1^2-c_2^2(a+bc_1^2)]\geq0$; and
$\gamma=a(a-\frac{c_1^2}{b})$ otherwise. With the exchanging $a\rightarrow b$ and $c_1\rightarrow c_2$,  we can obtain ``quantum $B$ discord" $D_2({\sigma_{B|A}})$.

Finally the R\'{e}nyi-2 mutual information for the two-mode Gaussian state Eq.(\ref{w1}) is given by \cite{ZZ1}
\begin{equation}\label{Q5}
I_2({\sigma_{AB}})=\frac{1}{2}\ln(\det \mathcal{A} )+\frac{1}{2}\ln(\det \mathcal{B} )- \frac{1}{2}\ln(\det \sigma_{AB}).
\end{equation}

\section{Schwinger effect of bosonic fields \label{tools}}
Schwinger effect of quantum vacuum has been clearly described by many authors \cite{wd32,wd31,wd20}. For the sake of self consistency, we present a brief review of the process.
Consider an electric field along the z-direction, whose gauge potential satisfies $E_{z}(t)=-\partial A_{z}(t)/\partial t$. For a constant electric field $E_{z}(t)=E_0$, the quadri-potential in the four dimensional Minkowski spacetime with the metric $(+,-,-,-)$ is given by $A_\mu=(0,0,0,-E_0t)$. A scalar particle with mass $m$ and charge $q$ coupled to this constant electric field has the equation of motion (Klein-Gordon equation),
\begin{equation}\label{w5}
[(\partial_\mu-ieA_\mu)(\partial^\mu-ieA^\mu)+m^2]\Psi(t,x)=0,
\end{equation}
where $\Psi(t,x)$ is the scalar field.

When $t^{\rm in}=-\infty$ and $t^{\rm out}=+\infty$, the Klein-Gordon equation has the asymptotic complete sets
of mode functions $\{\mu^{\rm in},\nu^{\rm in}\}$ and $\{\mu^{\rm out},\nu^{\rm out}\}$ respectively\cite{wd20}. Here the superscripts `in' and `out' mean the input and output of the bosonic field, corresponding to respectively the absence and presence of the electric field.
Quantizing scalar field $\Psi(x)$ in terms of the input and output mode functions respectively, one can obtain the Bogoliubov transformations between the input and output operators \cite{wd20,wd28,wd29},
\begin{equation}\label{w7}
a^{\rm in}_\mathbf{k}=\alpha_\mathbf{k}^*a^{\rm out}_\mathbf{k}-\beta^*_\mathbf{k}b^{{\rm out} \dag}_{\mathbf{k}},
\end{equation}
\begin{equation}\label{w8}
b^{\rm in}_\mathbf{k}=\alpha_\mathbf{k}^*b^{\rm out}_\mathbf{k}-\beta^*_\mathbf{k}a^{{\rm out}\dag}_{\mathbf{k}},
\end{equation}
where $a_\mathbf{k}$ and $b_\mathbf{k}$ are the annihilation operators of particles and antiparticles with momentum $\mathbf{k}$, and
\begin{equation}\label{alpha}
  \begin{array}{cc}
    \alpha_\mathbf{k}=\frac{\sqrt{2 \pi}}{\Gamma(-\eta)}e^{\frac{-i\pi(\eta+1)}{2}}, & \beta_\mathbf{k}=e^{-i\pi\eta},
  \end{array}
\end{equation}
with $\eta=-\frac{1}{2}-i\frac{\zeta}{2}$ and $\zeta=\frac{m^2+k_{\bot}}{eE_0}$ fulfilling $|\alpha_\mathbf{k}|^2-|\beta_\mathbf{k}|^2=1$. The transverse momentum $k_{\bot}$ is defined by $k_{\bot}^{2}=k_{x}^{2}+k_{y}^{2}$.
Through the Bogoliubov transformations, one can find that an input vacuum state after Schwinger effect becomes as \cite{wd20,wd28,wd29}
\begin{equation}\label{w9}
|0_\mathbf{k}, 0_{-\mathbf{k}}\rangle^{\rm in}=\frac{1}{\alpha_\mathbf{k}}\sum_{n=0}^{\infty}\left(\frac{\beta_\mathbf{k}^*}{\alpha_\mathbf{k}^*}\right)^n|n_\mathbf{k}, n_{-\mathbf{k}}\rangle^{\rm out},
\end{equation}
where $|n_\mathbf{k}\rangle$ and $|n_{-\mathbf{k}}\rangle$ represent respectively the number states of particle and antiparticle with momentum $\mathbf{k}$ and the minus refers to antiparticle.

Eq.(\ref{w9}) can be rewritten as $|0_\mathbf{k}, 0_{-\mathbf{k}}\rangle^{\rm in}=U_{\mathbf{k}}|0_\mathbf{k}, 0_{-\mathbf{k}}\rangle^{\rm out}$, where $U_{\mathbf{k}}=S_{\mathbf{k}}P_{\mathbf{k}}$ with the phase operator $P_{\mathbf{k}}$ and the two-mode squeezing operator $S_{\mathbf{k}}$ given by
\begin{equation}\label{pk}
 \nonumber   P_{\mathbf{k}}=\exp[i\theta_{\mathbf{k}}(a^{{\rm out}\dag}_{\mathbf{k}}a^{\rm out}_{\mathbf{k}}+b^{{\rm out}\dag}_{\mathbf{k}}b^{\rm out}_{\mathbf{k}}+1)],
\end{equation}
\begin{equation}\label{sk}
 \nonumber   S_{\mathbf{k}}=\exp[r_{\mathbf{k}}(a^{{\rm out}\dag}_{\mathbf{k}}b^{{\rm out}\dag}_{\mathbf{k}}e^{2i\vartheta_{\mathbf{k}}}-a^{\rm out}_{\mathbf{k}}b^{\rm out}_{\mathbf{k}}e^{-2i\vartheta_{\mathbf{k}}})],
\end{equation}
here the squeeze parameter $r_{\mathbf{k}}$, the squeeze angle $\vartheta_{\mathbf{k}}$, and the phase angle $\theta_{\mathbf{k}}$ are determined by $\alpha_{\mathbf{k}}=e^{-i\theta_{\mathbf{k}}}\cosh r_{\mathbf{k}}$ and
$\beta^{*}_{\mathbf{k}}=-e^{-i\theta_{\mathbf{k}}}(e^{2i\vartheta_{\mathbf{k}}}\sinh r_{\mathbf{k}}).$ Observing that $P_{\mathbf{k}}|0_\mathbf{k}, 0_{-\mathbf{k}}\rangle^{\rm out}=\exp(i\theta_{\mathbf{k}})|0_\mathbf{k}, 0_{-\mathbf{k}}\rangle^{\rm out}$, thus up to an overall phase factor, $U_{\mathbf{k}}$ is equivalent to $S_{\mathbf{k}}$. Further, by absorbing the phases into the Bosonic operators, i.e., performing transformations $a^{\rm out}_{\mathbf{k}}\rightarrow a^{\rm out}_{\mathbf{k}} e^{-i\vartheta_{\mathbf{k}}}$ and $b^{\rm out}_{\mathbf{k}}\rightarrow b^{\rm out}_{\mathbf{k}} e^{-i\vartheta_{\mathbf{k}}}$, which is equivalent to rotating the squeezing direction of the input fields by an angle $\vartheta_{\mathbf{k}}$, we have
\begin{equation}\label{sk1}
S_{\mathbf{k}}=\exp[r_{\mathbf{k}}(a^{{\rm out}\dag}_{\mathbf{k}}b^{{\rm out}\dag}_{\mathbf{k}}-a^{\rm out}_{\mathbf{k}}b^{\rm out}_{\mathbf{k}})].
\end{equation}
In the phase space defined in the above section, $S_{\mathbf{k}}$ may be written as
\begin{eqnarray}\label{w10}
 S_{\mathbf{k}}= \left(
                      \begin{array}{cc}
                        |\alpha_\mathbf{k}|I_{2} & |\beta_\mathbf{k}|Z_2 \\
                        |\beta_\mathbf{k}|Z_2 & |\alpha_\mathbf{k}|I_{2} \\
                      \end{array}
                    \right),
\end{eqnarray}
where $I_{2}$ is the unity matrix in $2\times 2$ space and $Z_2$ the third Pauli matrix. The deduction suggests that the influence of Schwinger effect on the bosonic fields is equivalent to a bosonic amplication channel described by the squeezing operator $S_{\mathbf{k}}$.

Having the transformation of Eq.(\ref{sk1}) or Eq.(\ref{w10}) in hand, one can study in principle the Schwinger effect of any quantum states, including Gaussian states (coherent state, squeezed state and thermal state), as well as Fock states and their superposition. As an exemplary example, we will study the Schwinger effect of a two-mode squeezed state shared by Alice and Bob, with the momentum of each mode given by $\mathbf{p}$ and $\mathbf{q}$ respectively. We will consider two cases: (A) unilateral Schwinger effect, i.e., only Bob's mode encounters the influence of a constant electric field; (B) bilateral Schwinger effect, i.e., both Alice's and Bob's modes encounter the influence of electric fields. For simplicity, we assume that the modes $\mathbf{p}$ and $\mathbf{q}$ for the bilateral Schwinger effect have the same magnitude and direction, and encounter the same constant electric fields, so that $p_\bot=q_\bot$ and $\alpha_\mathbf{p}=\alpha_\mathbf{q}\equiv\alpha$, $\beta_\mathbf{p}=\beta_\mathbf{q}\equiv\beta$.

\section{Schwinger effect of Gaussian correlations}
With the above preparations, we can now study the influence of Schwinger effect on the Gaussian correlations. The input state we considered is a two-mode squeezed vacuum state with CM \cite{L23,CV}
\begin{eqnarray}\label{w11}
 \sigma_{AB}(\mathbf{p},\mathbf{q})^{\rm in}= \left(\!\!\begin{array}{cc}
\cosh(2s)I_{2} & \sinh(2s)Z_{2}\\
\sinh(2s)Z_{2} & \cosh(2s)I_{2}
\end{array}\!\!\right),
\end{eqnarray}
where $s$ is the squeezing parameter, and $\mathbf{p}$ and $\mathbf{q}$ denote respectively the momenta of the two modes shared by Alice and Bob. Observing that the input modes of the antiparticles $-\mathbf{p}$ and $-\mathbf{q}$ are in vacuum states, Eq.(\ref{w11}) also can be rewritten as
\begin{equation}\label{w12}
\sigma_{AB}(\mathbf{p},\mathbf{q},-\mathbf{p},-\mathbf{q})^{\rm in}=\sigma_{AB}(\mathbf{p},\mathbf{q})^{\rm in}\oplus I(-\mathbf{p},-\mathbf{q}),
\end{equation}
where $I(-\mathbf{p},-\mathbf{q})$ denotes the CM of vacuum states. Now we go on our process in two cases.

\subsection{Unilateral Schwinger effect}
When only Bob's mode $\mathbf{q}$ suffers from a constant electric field, then the output state of the whole system becomes \cite{L23,CV}
\begin{eqnarray}\label{Q6}
\nonumber \sigma_{AB}(\mathbf{p},\mathbf{q},-\mathbf{p},-\mathbf{q})^{\rm out} &=& \big[I(\mathbf{p},-\mathbf{p}) \oplus  S(\mathbf{q},-\mathbf{q})] \big[\sigma_{AB}(\mathbf{p},\mathbf{q},-\mathbf{p},-\mathbf{\mathbf{q}})^{\rm in}\big]\\&& \big[I(\mathbf{p},-\mathbf{p}) \oplus  S(\mathbf{q},-\mathbf{q})\big]\,^{\sf T},
\end{eqnarray}
where operation $S(\mathbf{q},-\mathbf{q})$ is given by Eq.(\ref{w10}).

For unilateral Schwinger effect, we are interested in the correlation between modes of $(\mathbf{p}, \mathbf{q})$ and modes of $(\mathbf{p}, \mathbf{-q})$ for the output state.
By tracing over the field modes of $(\mathbf{-p}, \mathbf{-q})$ on Eq.(\ref{Q6}),
we obtain the CM for modes of $(\mathbf{p}, \mathbf{q})$ of the output fields as
\begin{equation}\label{Q7}
\sigma_{AB}(\mathbf{p},\mathbf{q})= \left( {\begin{array}{*{20}{c}}
\mathcal{A}_{\mathbf{p},\mathbf{q}} & \mathcal{X}_{\mathbf{p},\mathbf{q}}\\
{{\mathcal{X}_{\mathbf{p},\mathbf{q}}^{\sf T}}} & \mathcal{B}_{\mathbf{p},\mathbf{q}}\\
\end{array}} \right),
\end{equation}
where $\mathcal{A}_{\mathbf{p},\mathbf{q}}=[\cosh(2s) ]I_2$,
$\mathcal{B}_{\mathbf{p},\mathbf{q}}=[\cosh(2s) |\alpha|^2 + |\beta|^2]I_2$
and $\mathcal{X}_{\mathbf{p},\mathbf{q}}=[|\alpha|\sinh(2s)]Z_2$, with $\alpha$ and $\beta$ given by Eq.(\ref{alpha}). For brevity, we have omitted the superscript `out'.

If tracing over the field modes $(\mathbf{-p}, \mathbf{q})$, we obtain CM for field modes $(\mathbf{p}, \mathbf{-q})$ as
\begin{equation}\label{Q11}
\sigma_{AB}(\mathbf{p},\mathbf{-q})= \left( {\begin{array}{*{20}{c}}
\mathcal{A}_{\mathbf{p},\mathbf{-q}} & \mathcal{X}_{\mathbf{p},\mathbf{-q}}\\
{{\mathcal{X}_{\mathbf{p},\mathbf{-q}}^{\sf T}}} & \mathcal{B}_{\mathbf{p},\mathbf{-q}}\\
\end{array}} \right),
\end{equation}
where $\mathcal{A}_{\mathbf{p},\mathbf{-q}}=[\cosh(2s) ]I_2$,
$\mathcal{B}_{\mathbf{p},\mathbf{-q}}=[|\alpha|^2 + \cosh(2s) |\beta|^2]I_2$
and $\mathcal{X}_{\mathbf{p},\mathbf{-q}}=[|\beta|\sinh(2s)]I_2$.

If we use the first set of measures, according to Eq.(\ref{w2})-(\ref{w3}), we have the logarithmic negativity, von-Neumann discord and von-Neumann mutual information,
\begin{eqnarray}\label{QQ8}
\nonumber N_{1}(\sigma_{AB}(\mathbf{p},\mathbf{q}))&=&\frac{1}{2}-\frac{1}{2}\log\{\cosh^2(2s)+2|\alpha|^2\sinh^2(2s)+\zeta^{2}\\
&-&2[\cosh(2s)+\zeta]
\sqrt{\cosh^4(s)|\beta|^4+|\alpha|^2\sinh^2(2s)}\},
\end{eqnarray}
\begin{equation}\label{QQ9}
D_{1}(\sigma_{A|B}(\mathbf{p},\mathbf{q}))=f(\zeta)-f(\eta),
\end{equation}
\begin{equation}\label{QQ10}
D_{1}(\sigma_{B|A}(\mathbf{p},\mathbf{q}))=f[ \cosh(2s)]+f(1+2|\beta|^{2})-f(\eta),
\end{equation}
\begin{equation}\label{QQ11}
I_{1}(\sigma_{AB}(\mathbf{p},\mathbf{q}))=f[\cosh(2s) ]+f(\zeta)-f(\eta),
\end{equation}
for the correlations between modes $(\mathbf{p}, \mathbf{q})$, and
\begin{eqnarray}\label{QQ12}
N_{1}(\sigma_{AB}(\mathbf{p},\mathbf{-q}))=0,
\end{eqnarray}
\begin{equation}\label{QQ13}
D_{1}(\sigma_{A|B}(\mathbf{p},\mathbf{-q}))=f(\xi)-f({\nu}_{-})-f(\nu_{+})+f(\varepsilon_{1}),
\end{equation}
\begin{equation}\label{QQ14}
D_{1}(\sigma_{B|A}(\mathbf{p},\mathbf{-q}))=f[\cosh(2s)]-f({\nu}_{-})-f({\nu}_{+})+f(\varepsilon_{2}),
\end{equation}
\begin{equation}\label{QQ15}
I_{1}(\sigma_{AB}(\mathbf{p},\mathbf{-q}))=f[\cosh(2s) ]-f({\nu}_{-})-f({\nu}_{+})+f(\xi),
\end{equation}
for the correlations between modes $(\mathbf{p}, \mathbf{-q})$.
Where $\zeta=\cosh(2s)|\alpha|^2 + |\beta|^2$, ${\eta}=1+2|\beta|^2\cosh(s)$,
$\xi=|\alpha|^{2}+|\beta|^{2}\cosh(2s)$,  $\varepsilon_{1}=\frac{|\beta|^{2}+(|\beta|^{2}+2)\cosh(2s)}{2+[1+\cosh(2s)]|\beta|^{2}}$, $\varepsilon_{2}=2|\beta|^2+1$, and
$$
{\nu}_\mp=\frac{1}{\sqrt{2}}\left\{\cosh^2(2s)+\xi^2-2|\beta|^2\sinh^2(2s)\mp\sqrt{[\cosh^2(2s)+\xi^2-2|\beta|^2\sinh^2(2s)]^2-4\zeta}\right\}^{\frac{1}{2}}.
$$
These correlations in Eqs.(\ref{QQ8})-(\ref{QQ15}) depend on both the strength of the electric field $E_0$ and the squeezing parameter $s$.
When $E_0=0$, they reduce to their input values.

\begin{figure}
\begin{minipage}[t]{0.5\linewidth}
\centering
\includegraphics[width=3.0in,height=5.2cm]{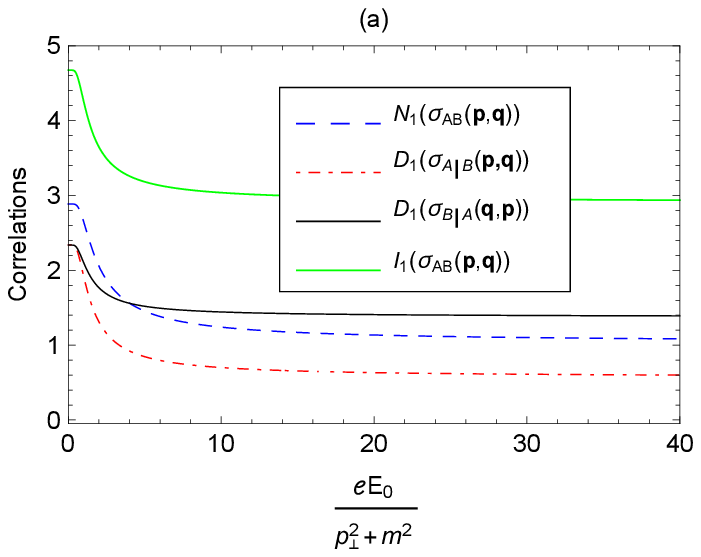}
\label{fig1a}
\end{minipage}%
\begin{minipage}[t]{0.5\linewidth}
\centering
\includegraphics[width=3.0in,height=5.2cm]{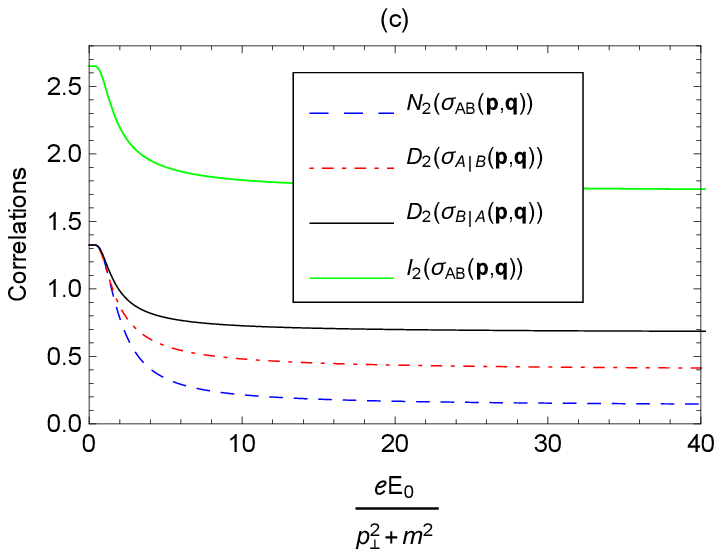}
\label{fig1c}
\end{minipage}%

\begin{minipage}[t]{0.5\linewidth}
\centering
\includegraphics[width=3.0in,height=5.2cm]{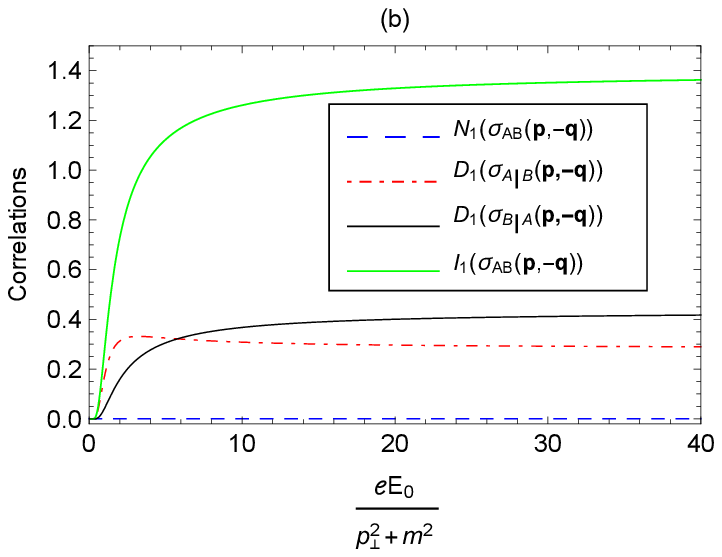}
\label{fig1b}
\end{minipage}%
\begin{minipage}[t]{0.5\linewidth}
\centering
\includegraphics[width=3.0in,height=5.2cm]{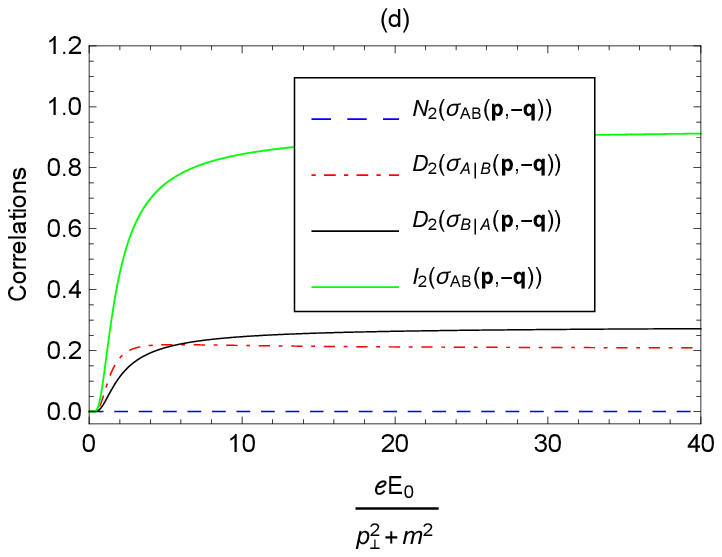}
\label{fig1d}
\end{minipage}%
\caption{Entanglement, discord, and mutual information as functions of the dimensionless parameters $\frac{eE_0}{p_\bot^2+m^2}$ for the unilateral Schwinger effect, where the squeezing parameter $s=1$. (a) and (b) for the first set of measure of correlations, and (c) and (d) for the second one.}
\label{Fig1}
\end{figure}

In Fig.\ref{Fig1}(a) and (b), we plot the correlations as functions of the dimensionless electric field $\frac{eE_0}{p_\bot^2+m^2}$, for fixed squeezing parameter $s=1$. We see that all the correlations between modes $(\mathbf{p}, \mathbf{q})$ decrease monotonically with the increase of $E_0$, but have nonzero asymptotic values when $E_0\rightarrow \infty$.
At the same time, the correlations between modes $(\mathbf{p}, \mathbf{-q})$, except for the entanglement $N_{1}(\sigma_{AB}(\mathbf{p},\mathbf{-q}))$, increase from zero, and reach bounded asymptotic values when $E_0\rightarrow \infty$.
A point needed to emphasize is that the unilateral Schwinger effect can produce asymmetry of quantum discord, i.e., $D(\sigma_{A|B}(\mathbf{p},\mathbf{q}))\neq D(\sigma_{B|A}(\mathbf{p},\mathbf{q}))$ and $D(\sigma_{A|B}(\mathbf{p},\mathbf{-q}))\neq D(\sigma_{B|A}(\mathbf{p},\mathbf{-q}))$ in general.

For the second set of measures, correlations may be calculated in principle through Eq.(\ref{Q3})-(\ref{Q5}). However they are too tedious and here we only give the analytical expressions of the mutual information, and show the entanglement and discord numerically.
The R\'{e}nyi-2 entropy mutual information between modes $(\mathbf{p}, \mathbf{q})$ or modes $(\mathbf{p}, \mathbf{-q})$ reads respectively,
\begin{equation}\label{Q15}
I_2(\sigma_{AB}(\mathbf{p},\mathbf{q}))=\ln[\frac{\zeta\cosh(2s)}{\xi}],
\end{equation}
\begin{equation}\label{Q16}
I_2(\sigma_{AB}(\mathbf{p},\mathbf{-q}))=\ln[\frac{\xi\cosh(2s)}{\zeta}],
\end{equation}
with $\zeta$ and $\xi$ as given previously.

In Fig.\ref{Fig1}(c) and (d), we plot the R\'{e}nyi-2 entropy correlations between modes $(\mathbf{p}, \mathbf{q})$ and between modes $(\mathbf{p}, \mathbf{-q})$ as functions of the dimensionless electric field $\frac{eE_0}{p_\bot^2+m^2}$, for fixed squeezing parameter $s=1$.
We see that the correlation behaviors for this case are very similar to that for the first set of measures.

It is worthwhile to point out that the authors in reference \cite{wd31} studied a model in which a free qubit couples to a bosonic mode with Schwinger effect. They studied, by the first set of measures, the Schwinger effect to the influence of entanglement and mutual information. We find that the entanglement and mutual information's behaviors presented by Fig.\ref{Fig1} are very similar to the one of reference \cite{wd31}, though a free qubit is now replaced by the continuous-variable bosonic mode.

\begin{figure}
\begin{minipage}[t]{0.5\linewidth}
\centering
\includegraphics[width=3.0in,height=5.2cm]{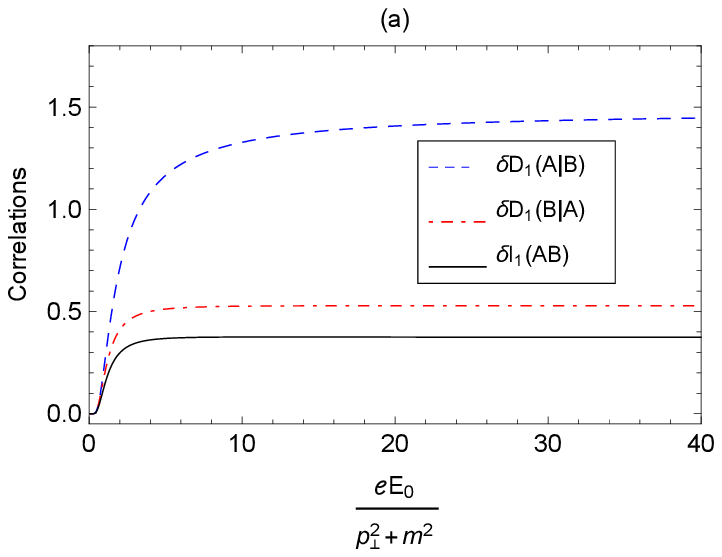}
\label{fig1a}
\end{minipage}%
\begin{minipage}[t]{0.5\linewidth}
\centering
\includegraphics[width=3.0in,height=5.2cm]{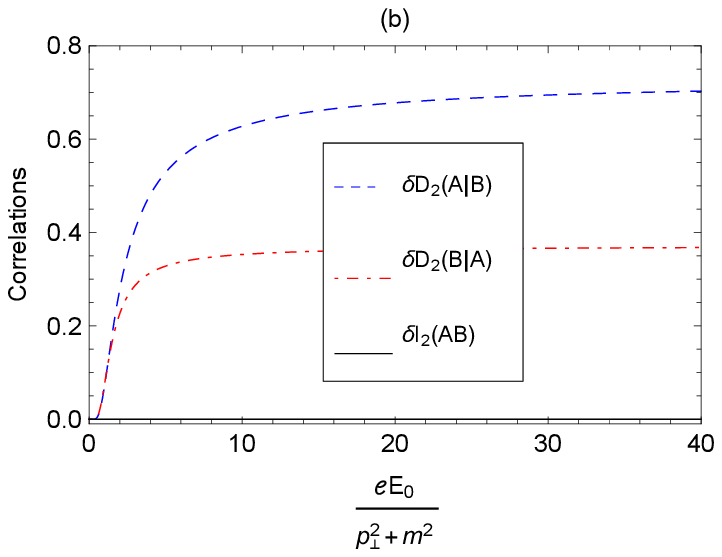}
\label{fig1c}
\end{minipage}%
\caption{Monogamy scores of discord and mutual information as functions of the dimensionless parameters $\frac{eE_0}{p_\bot^2+m^2}$ for the unilateral Schwinger effect, where the squeezing parameter $s=1$. (a) and (b) are respectively corresponding to the first and the second sets of measure.}
\label{Fig2}
\end{figure}

The monogamy or redistribution of correlations is an interesting problem in the dynamics of correlations. To study the problem, we define the monogamy score of the correlation $R$ for the case of unilateral Schwinger effect as
\begin{equation}\label{Q23}
\delta R_{j}(AB)=R_{j}(\sigma_{AB}(\mathbf{p},\mathbf{q})^{\rm in})-R_{j}(\sigma_{AB}(\mathbf{p},\mathbf{q})^{\rm out})
-R_{j}(\sigma_{AB}(\mathbf{p},\mathbf{-q})^{\rm out}).
\end{equation}
This quantity describes the change or/and redistribution of the correlation $R$. If $\delta R_{j}(AB)=0$, we say that the correlation $R$ is conservative and part of which is redistributed form $(\mathbf{p},\mathbf{q})$ to $(\mathbf{p},\mathbf{-q})$ through the Schwinger effect. Otherwise the correlation is said to be nonconservative. In this definition, the quantity $R$ may be the $A$-discord  $D(A|B)$, $B$-discord $ D(B|A)$, or the mutual information $I(AB)$, and the subscript $j=1,2$ denotes the first or the second set of measure of correlations. For clarity, we recover the superscripts to clearly mark the input and output states.

From Fig.\ref{Fig1}, we can see already that the discord and mutual information for the both sets of measures can be partially redistributed from modes $(\mathbf{p},\mathbf{q})$ to modes $(\mathbf{p},\mathbf{-q})$. However, the entanglement obviously can not be redistributed, because the entanglement between modes $(\mathbf{p},\mathbf{-q})$ remains always zero.

Fig.\ref{Fig2} shows the evolution of the correlation monogamy versus the dimensionless electric field. We see that the discord for both sets of measure is nonconservative and is lossy, but the mutual information has different behaviour: The von-Neumann mutual information is lossy, while the R\'{e}nyi-2 entropy mutual information is conservative. This result contrasts with that of unilateral Schwinger effect of qubit-boson system\cite{wd31}, where the von-Neumann mutual information is conservative. The conservativeness of the R\'{e}nyi-2 entropy mutual information also can be verified analytically. According to Eqs.(\ref{Q15})-(\ref{Q16}), one easily obtains
\begin{equation}\label{vv1}
I_2(\sigma_{AB}(\mathbf{p},\mathbf{q}))+I_2(\sigma_{AB}(\mathbf{p},\mathbf{-q}))=2\ln[\cosh(2s)],
\end{equation}
with $2\ln[\cosh(2s)]$ is the initially mutual information for Eq.(\ref{w11}).

\subsection{Bilateral Schwinger effect}

When both Alice's mode $\mathbf{p}$ and Bob' mode $\mathbf{q}$ are influenced by the same electric fields, the output state of the whole system is \cite{L23,CV}
\begin{eqnarray}\label{w13}
\nonumber \sigma_{AB}(\mathbf{p},\mathbf{q},-\mathbf{p},-\mathbf{q})^{\rm out} &=& \big[S(\mathbf{p},-\mathbf{p}) \oplus  S(\mathbf{q},-\mathbf{q})] \big[\sigma_{AB}(\mathbf{p},\mathbf{q},-\mathbf{p},-\mathbf{\mathbf{q}})^{\rm in}\big]\\&& \big[S(\mathbf{p},-\mathbf{p}) \oplus  S(\mathbf{q},-\mathbf{q})\big]\,^{\sf T}.
\end{eqnarray}

Now we are interested in four sets of correlations, $(\mathbf{p}, \mathbf{q})$, $(\mathbf{p}, \mathbf{-q})$, $(\mathbf{-p}, \mathbf{q})$ and $(\mathbf{-p}, \mathbf{-q})$. Considering the symmetry of the problem, we need only to calculate the three sets of correlations, $(\mathbf{p}, \mathbf{q})$, $(\mathbf{p}, \mathbf{-q})$, and $(\mathbf{-p}, \mathbf{-q})$. The corresponding reduced CM are respectively
\begin{equation}\label{w14}
\sigma_{AB}(\mathbf{p},\mathbf{q})= \left( {\begin{array}{*{20}{c}}
\mathcal{A}_{\mathbf{p},\mathbf{q}} & \mathcal{X}_{\mathbf{p},\mathbf{q}}\\
{{\mathcal{X}_{\mathbf{p},\mathbf{q}}^{\sf T}}} & \mathcal{B}_{\mathbf{p},\mathbf{q}}\\
\end{array}} \right),
\end{equation}
\begin{equation}\label{w18}
\sigma_{AB}(\mathbf{p},\mathbf{-q})= \left( {\begin{array}{*{20}{c}}
\mathcal{A}_{\mathbf{p},\mathbf{-q}} & \mathcal{X}_{\mathbf{p},\mathbf{-q}}\\
{{\mathcal{X}_{\mathbf{p},\mathbf{-q}}^{\sf T}}} & \mathcal{B}_{\mathbf{p},\mathbf{-q}}\\
\end{array}} \right),
\end{equation}
\begin{equation}\label{w22}
\sigma_{AB}(\mathbf{-p},\mathbf{-q})= \left( {\begin{array}{*{20}{c}}
\mathcal{A}_{\mathbf{-p},\mathbf{-q}} & \mathcal{X}_{\mathbf{-p},\mathbf{-q}}\\
{{\mathcal{X}_{\mathbf{-p},\mathbf{-q}}^{\sf T}}} & \mathcal{B}_{\mathbf{-p},\mathbf{-q}}\\
\end{array}} \right),
\end{equation}
where $\mathcal{A}_{\mathbf{p},\mathbf{q}}=\mathcal{B}_{\mathbf{p},\mathbf{q}}=\mathcal{A}_{\mathbf{p},\mathbf{-q}}=[\cosh(2s) |\alpha|^2 + |\beta|^2]I_2$, $\mathcal{B}_{\mathbf{p},\mathbf{-q}}=\mathcal{A}_{\mathbf{-p},\mathbf{-q}}=\mathcal{B}_{\mathbf{-p},\mathbf{-q}}=[|\alpha|^2 + \cosh(2s) |\beta|^2]I_2$,
$\mathcal{X}_{\mathbf{p},\mathbf{q}}=[|\alpha|^2\sinh(2s)]Z_2$,
$\mathcal{X}_{\mathbf{p},\mathbf{-q}}=[|\alpha||\beta|\sinh(2s)]I_2$,
and $\mathcal{X}_{\mathbf{-p},\mathbf{-q}}=[|\beta|^2\sinh(2s)]Z_2$.

For the first set of measure, we calculate the logarithmic negativity, discord and mutual information as,
\begin{eqnarray}\label{w15}
N_{1}(\sigma_{AB}(\mathbf{p},\mathbf{q}))
= -\log[\zeta-\sinh(2s)|\alpha|^2],
\end{eqnarray}
\begin{eqnarray}\label{w16}
D_{1}(\sigma_{A|B}(\mathbf{p},\mathbf{q}))=D_{1}(\sigma_{B|A}(\mathbf{p},\mathbf{q}))=f(\zeta)+f(|\alpha|^2+|\beta|^2)-2f(\tau),
\end{eqnarray}
\begin{equation}\label{w17}
I_{1}(\sigma_{AB}(\mathbf{p},\mathbf{q}))=2f(\zeta)-2f(\tau),
\end{equation}
for the correlation between modes $(\mathbf{p}, \mathbf{q})$, and
\begin{eqnarray}\label{w19}
N_{1}(\sigma_{AB}(\mathbf{p},\mathbf{-q}))&=&\frac{1}{2}-\frac{1}{2}\log\{[1+\cosh^2(2s)](|\alpha|^4+|\beta|^4)\\
\nonumber &+&[\cosh(4s)+4\cosh(2s)-1]|\alpha|^2|\beta|^2\\
\nonumber &-&2\cosh^2(s)\sinh(s)(|\alpha|^2+|\beta|^2)\\
\nonumber &\times&\sqrt{\chi-3+\cosh(2s)(1+\chi)}\},
\end{eqnarray}
\begin{equation}\label{w20}
D_{1}(\sigma_{A|B}(\mathbf{p},\mathbf{-q}))=f(\xi)-f({\nu}'_-)
-f({\nu}'_+)+f(\varepsilon_{3}),
\end{equation}
\begin{equation}\label{ww20}
D_{1}(\sigma_{B|A}(\mathbf{p},\mathbf{-q}))=f(\zeta)-f({\nu}'_-)
-f({\nu}'_+)+f(\varepsilon_{2}),
\end{equation}
\begin{equation}\label{w21}
I_{1}(\sigma_{AB}(\mathbf{p},\mathbf{-q}))=f(\zeta)+f(\xi)-f({\nu}'_-)
-f({\nu}'_+),
\end{equation}
for the correlation between modes $(\mathbf{p}, \mathbf{-q})$, and
\begin{eqnarray}\label{w23}
N_{1}(\sigma_{AB}(\mathbf{-p},\mathbf{-q}))
=-\log[\xi-\sinh(2s)|\beta|^2],
\end{eqnarray}
\begin{eqnarray}\label{w24}
D_{1}(\sigma_{A|B}(\mathbf{-p},\mathbf{-q}))=D_{1}(\sigma_{B|A}(\mathbf{-p},\mathbf{-q}))=f(\xi)+f(\varepsilon_{4})-2f(\tau),
\end{eqnarray}
\begin{eqnarray}\label{w25}
I_{1}(\sigma_{AB}(\mathbf{-p},\mathbf{-q}))
=2f(\xi)
-2f(\tau),
\end{eqnarray}
for the correlation between modes $(\mathbf{-p}, \mathbf{-q})$.
Where $\zeta$, $\xi$, $\varepsilon_{2}$ are defined as before, and $\chi=(|\alpha|^2 + |\beta|^2)^2+4|\alpha|^4 |\beta|^4$, $\tau=\frac{1}{2}\sqrt{3+8\cosh(2s) |\alpha|^2|\beta|^2+\chi}$,
$$
{\nu}'_\mp=\frac{1}{2}\left\{2+\cosh(4s)+\chi+8\cosh(2s)|\alpha|^2|\beta|^2 \mp4\sqrt{2}\cosh(s)\sinh^2(s)\sqrt{\cosh(2s)+\chi}\right\}^{\frac{1}{2}},
$$
$$
\varepsilon_{3}=\frac{\cosh^2(2s)|\alpha|^2|\beta|^2+\{1-\frac{1}{2}[\cosh(4s)-3]|\alpha|^2\}|\beta|^2
+\cosh(2s)(|\alpha|^2+|\alpha|^4+|\beta|^4)}{1+\xi},
$$
$$\varepsilon_{4}=\frac{|\alpha|^4+|\beta|^4+2\cosh(2s)|\alpha|^{2} |\beta|^2+\xi}{1+\xi}.$$

\begin{figure}
\begin{minipage}[t]{0.5\linewidth}
\centering
\includegraphics[width=3.0in,height=5.2cm]{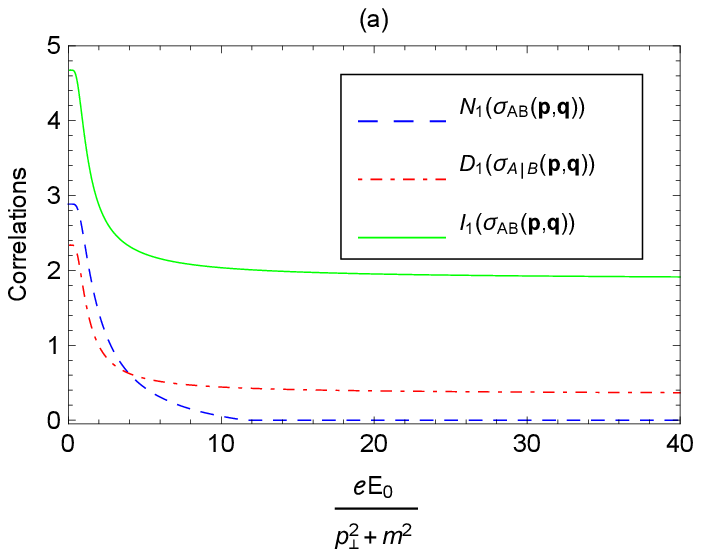}
\label{fig1a}
\end{minipage}%
\begin{minipage}[t]{0.5\linewidth}
\centering
\includegraphics[width=3.0in,height=5.2cm]{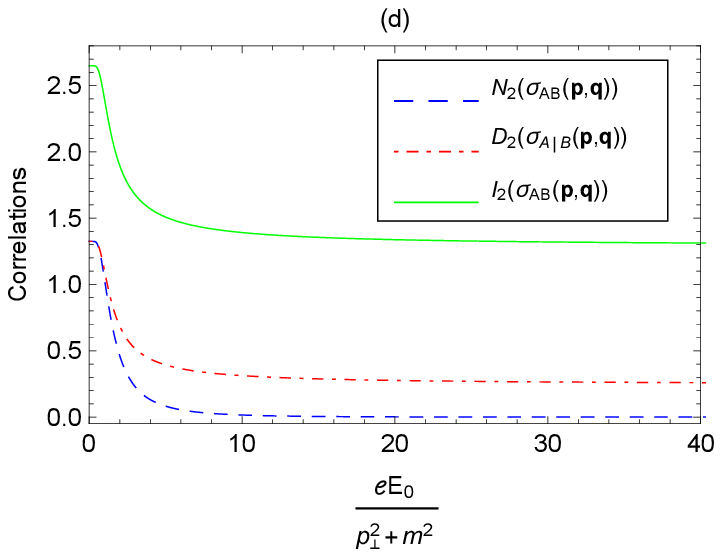}
\label{fig1c}
\end{minipage}%

\begin{minipage}[t]{0.5\linewidth}
\centering
\includegraphics[width=3.0in,height=5.2cm]{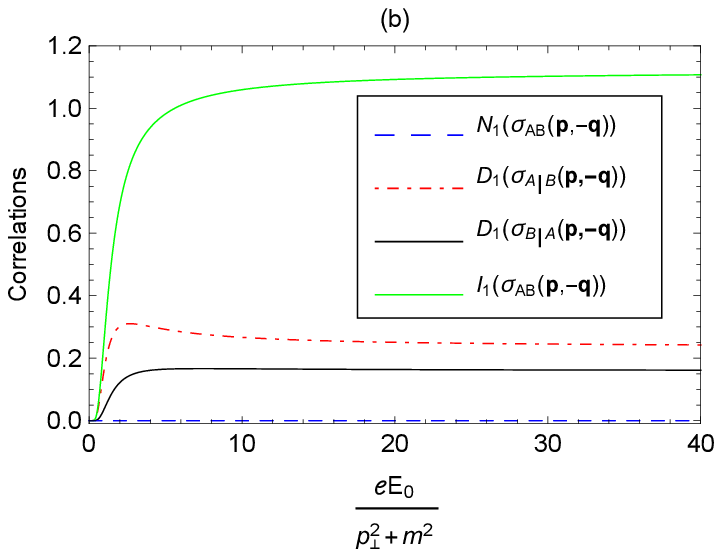}
\label{fig1b}
\end{minipage}%
\begin{minipage}[t]{0.5\linewidth}
\centering
\includegraphics[width=3.0in,height=5.2cm]{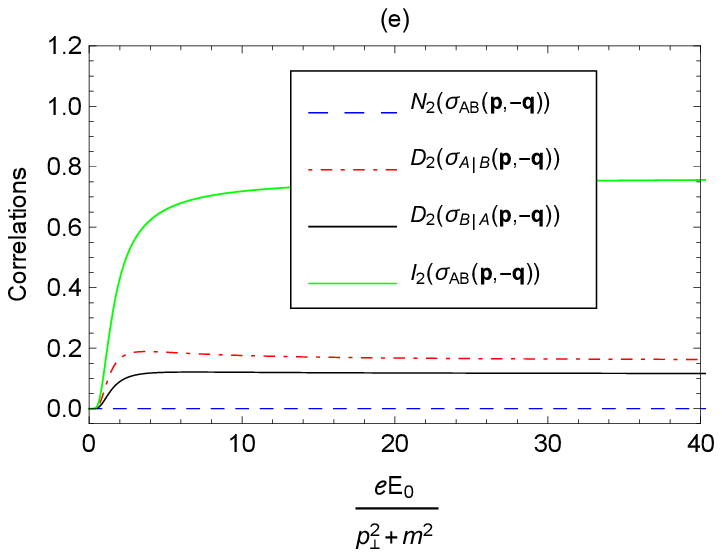}
\label{fig1d}
\end{minipage}%

\begin{minipage}[t]{0.5\linewidth}
\centering
\includegraphics[width=3.0in,height=5.2cm]{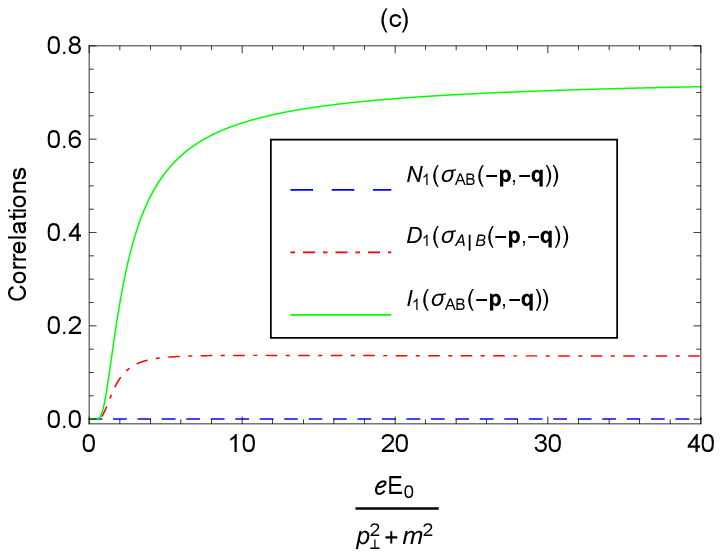}
\label{fig1b}
\end{minipage}%
\begin{minipage}[t]{0.5\linewidth}
\centering
\includegraphics[width=3.0in,height=5.2cm]{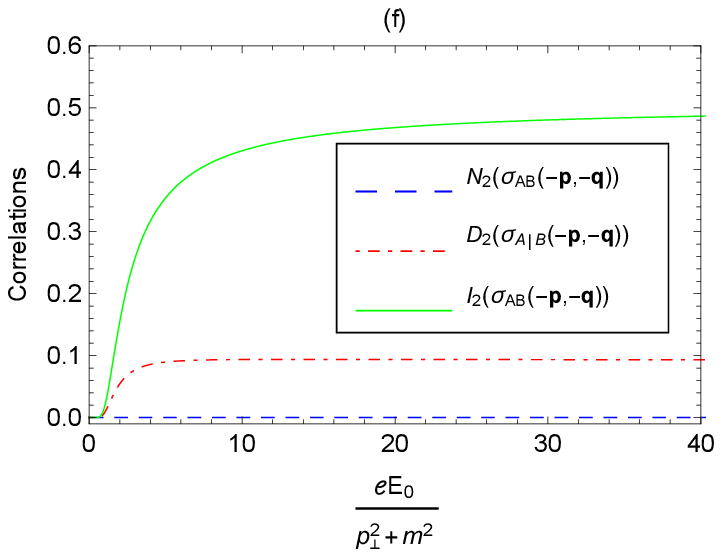}
\label{fig1d}
\end{minipage}%

\caption{Entanglement, discord, and mutual information as functions of the dimensionless parameters $\frac{eE_0}{p_\bot^2+m^2}$ for the bilateral Schwinger effect, where the squeezing parameter $s=1$. (a)-(c) for the first set of measure of correlations, and (d)-(f) for the second one. }
\label{Fig3}
\end{figure}

For the second set of measure, the R\'{e}nyi-2 entropy mutual information between the interesting modes read respectively,
\begin{equation}\label{Q17}
I_2(\sigma_{AB}(\mathbf{p},\mathbf{q})=\ln\frac{\zeta^2}{\xi^{2}+|\beta|^4\sinh^{2}(2s)},
\end{equation}
\begin{equation}\label{Q18}
I_2(\sigma_{AB}(\mathbf{p},\mathbf{-q}))=\ln\frac{\zeta\xi}{|\alpha|^2|\zeta+|\beta|^2\xi}.
\end{equation}
\begin{equation}\label{Q19}
I_2(\sigma_{AB}(\mathbf{-p},\mathbf{-q}))=\ln\frac{\xi^{2}}{\xi^{2}+|\beta|^4\sinh^{2}(2s)}.
\end{equation}

In Fig.\ref{Fig3}, we plot the evolution of the various correlations versus the dimensionless electric field for the case of bilateral Schwinger effect. It is shown that in general the behaviours are similar to that of the unilateral Schwinger effect: The correlation between modes $(\mathbf{p},\mathbf{q})$ degrades monotonically, and at the same time the correlation (discord and mutual information) between modes $(\mathbf{p},\mathbf{-q})$, $(\mathbf{-p},\mathbf{q})$ and $(\mathbf{-p},\mathbf{-q})$ produces. In other words, except for entanglement, discord and mutual information can be redistributed partially. The difference is that the entanglement of both $N_{1}(\sigma_{AB}(\mathbf{p},\mathbf{q}))$ and $N_{2}(\sigma_{AB}(\mathbf{p},\mathbf{q}))$ in the case of bilateral Schwinger effect takes place sudden death. The position of sudden death of $N_{1}(\sigma_{AB}(\mathbf{p},\mathbf{q}))$ fulfils $$\frac{eE_0}{p_\bot^2+m^2}=\frac{\pi}{\ln[\frac{1+\cosh(2s)-\sinh(2s)}{1-\cosh(2s)+\sinh(2s)}]}.$$ Except for entanglement, the discord and mutual information in Fig.\ref{fig3} decay to nonzero asymptotical values or increase to bounded asymptotical values when $E_0\rightarrow \infty$.

Note that the authors in \cite{wd32} studied the bilateral Schwinger effect of fermion-fermion fields by use of the first set of measures. We find three differences between the two settings: Firstly, in the degradation of entanglement between fermion modes $(\mathbf{p},\mathbf{q})$, it would not take place entanglement sudden death, or even has nonzero asymptotical value when $E_0\rightarrow \infty$. Secondly, the entanglement between fermion modes $(\mathbf{p},\mathbf{-q})$ and $(\mathbf{-p},\mathbf{q})$ can be produced. Lastly, the entanglement and mutual information between fermion modes $(\mathbf{p},\mathbf{-q})$ and $(\mathbf{-p},\mathbf{q})$ change non-monotonically with the increase of electric field. These behaviours are contrasted with the corresponding bosonic setting.

\begin{figure}
\begin{minipage}[t]{0.5\linewidth}
\centering
\includegraphics[width=3.0in,height=5.2cm]{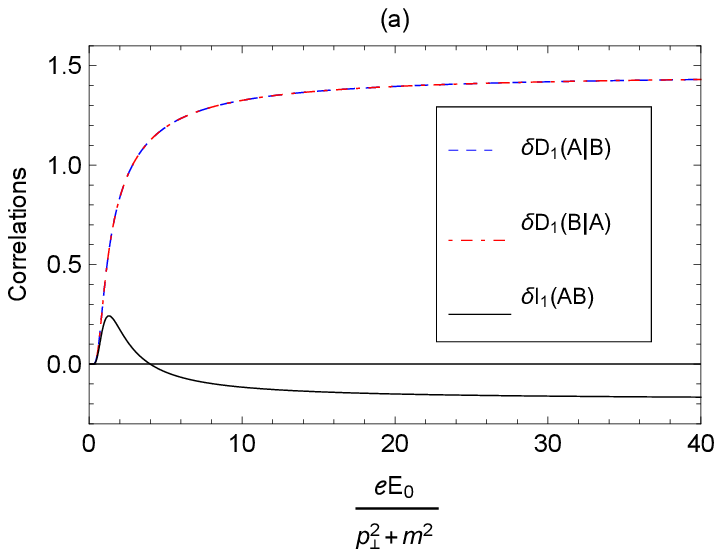}
\label{fig1a}
\end{minipage}%
\begin{minipage}[t]{0.5\linewidth}
\centering
\includegraphics[width=3.0in,height=5.2cm]{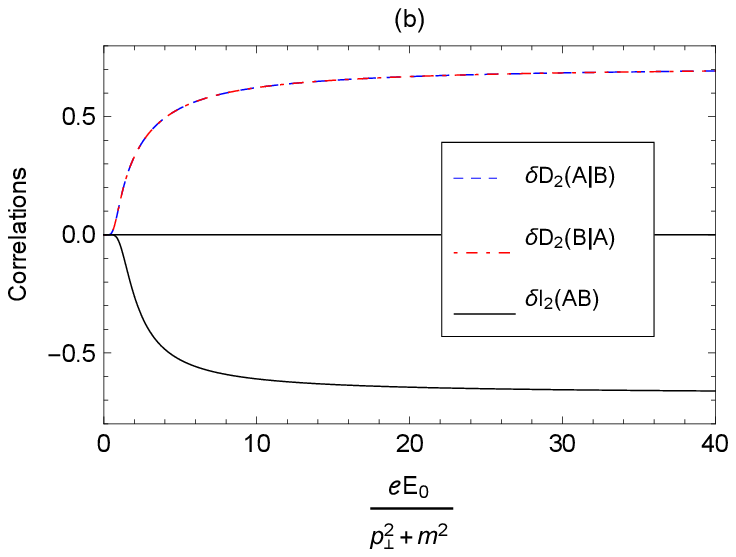}
\label{fig1c}
\end{minipage}%
\caption{Monogamy scores of discord and mutual information as functions of the dimensionless parameters $\frac{eE_0}{p_\bot^2+m^2}$ for the bilateral Schwinger effect, where the squeezing parameter $s=1$. (a) and (b) are respectively corresponding to the first and the second sets of measure.}
\label{Fig4}
\end{figure}

For bilateral Schwinger effect, we now define the monogamy scores of correlations
\begin{eqnarray}\label{BQ23}
\nonumber \delta R_{j}(AB)&=&R_{j}(\sigma_{AB}(\mathbf{p},\mathbf{q})^{\rm in})-R_{j}(\sigma_{AB}(\mathbf{p},\mathbf{q})^{\rm out})
-R_{j}(\sigma_{AB}(\mathbf{p},\mathbf{-q})^{\rm out})\\
&-&R_{j}(\sigma_{AB}(\mathbf{-p},\mathbf{q})^{\rm out})-R_{j}(\sigma_{AB}(\mathbf{-p},\mathbf{-q})^{\rm out}),
\end{eqnarray}
here the quantity $R$ may be the $A$-discord  $D(A|B)$, $B$-discord $ D(B|A)$, or the mutual information $I(AB)$, and the subscript $j=1,2$ denotes the first or the second set of measure of correlations.
In Fig.\ref{Fig4}, we plot these monogamy scores as functions of the dimensionless electric field $\frac{eE_0}{p_\bot^2+m^2}$. It is shown that there is no conservative quantity of correlations for the case of bilateral Schwinger effect.
Like in case of unilateral Schwinger effect, the entanglement and discord are always lossy [$\delta N>0$, $\delta D>0$]. Different from the case of unilateral Schwinger effect, now we find that $\delta I_{2}(A|B)<0$ and  $\delta I_{1}(A|B)$ is also negative when $E_{0}$ is larger than certain critical value. This is to say the bilateral Schwinger effect can produce new (besides redistribution) mutual information between Alice and Bob, in sharp contrast with the case of unilateral Schwinger effect.
As the discord is lossy, the production of mutual information means actually the production of classical correlation between Alice and Bob.


 \section{Conclusions and discussions}

We have studied the Schwinger effect of quantum entanglement, discord and mutual information of continuous-variable two-mode squeezed states in constant electric fields. In the process of analyses, we have considered the two cases of unilateral and bilateral Schwinger effect, and have employed two sets of measures. We have found that macroscopically the behaviours are the same: All the correlations between the original modes of particle-particle degrade monotonically, at the same time the correlations (except for entanglement) between the modes of particle-antiparticle or antiparticle-antiparticle increase from zero to bounded asymptotic values, with the increase of Schwinger electric fields. However microscopically there are differences: The entanglement between the original modes of particle-particle decays to a nonzero asymptotic value for the case of unilateral Schwinger effect, but takes place sudden death for the case of bilateral Schwinger effect.

We have also studied the problem of monogamy of correlations. We found that entanglement, whether measured by Logarithmic negativity or based on R\'{e}nyi-2 entropy, for both unilateral and bilateral Schwinger effects, can not be redistributed, while discord and mutual information always can be redistributed in the considered situations. In the process of redistribution, the total quantity of discord is always degraded, but the mutual information has different behavior: The total quantity of mutual information is non-increasing (not always decreasing) for the unilateral Schwinger effect and can be produced for the bilateral Schwinger effect. An interesting result is that the R\'{e}nyi-2 entropy mutual information for the case of unilateral Schwinger effect is conservative.

The Schwinger effect of correlations of boson-boson fields has some different features compared with that of fermion-fermion fields. For example, in the case of bilateral Schwinger effect of fermion fields \cite{wd32}, there is no sudden death in the degradation of entanglement between the original fermion modes of particle-particle, and the entanglement between fermion modes of particle-antiparticle and antiparticle-antiparticle can be produced. Further, the entanglement and mutual information between the fermion modes of particle-antiparticle change non-monotonically with the increase of electric field. All these behaviours are in contrast with the results of the boson-boson Schwinger effect presented in this paper. The reason that leads to these differences originates perhaps from the difference in statistics, which leads to the Bogoliubov coefficients in Eqs.(\ref{w7})-(\ref{w8}) fulfilling $|\alpha_\mathbf{k}|^2-|\beta_\mathbf{k}|^2=1$ for bosonic fields and $|\alpha_\mathbf{k}|^2+|\beta_\mathbf{k}|^2=1$ for fermion fields \cite{wd20}.

There are also some differences for the Schwinger effects of correlation between boson-boson system and the qubit-boson system \cite{wd31}. For example, in the case of unilateral Schwinger effect of qubit-boson field, the mutual information based on von-Neumann entropy is conservative. However, for the corresponding boson-boson system, we have found instead the conservativeness of the R\'{e}nyi-2 entropy mutual information.

Finally, we would like to compare the Schwinger effect of correlation induced by electric fields with the Unruh effect of correlation induced by accelerations \cite{L23}. Both the two effects involve the production of particle-antiparticle pairs, and thus have very similar behaviors on the quantum correlations. However, there are still some differences. For example, in the degradation of correlations, the R\'{e}nyi-2 entropy discord between modes of particle-particle asymptotically vanishes in the infinite acceleration limit, but approaches to a nonzero value in the limit of infinite electric field. The same result is also valid for the degradation of the R\'{e}nyi-2 entropy entanglement in the cases of unilateral Unruh (unilateral Schwinger) effect. Mathematically, these differences is related to the relation  $|\alpha_\mathbf{k}|^2-|\beta_\mathbf{k}|^2=1$, in which $|\beta_\mathbf{k}|$ is less than one for Schwinger effect and arbitrary for Unruh effect.

\begin{acknowledgments}
This work is supported by the National Natural
Science Foundation of China (Grant No. 11275064), and the Construct Program of the National Key Discipline.	
\end{acknowledgments}


\end{document}